\font\bbb=msbm10                                                   

\def\C{\hbox{\bbb C}}

\def\Z{\hbox{\bbb Z}}

\def\APS{{\sl Acta Physica Slovaca}}

\def\CMP{{\sl Commun.\ Math.\ Phys.}}
\def\CPAM{{\sl Commun.\ Pure Appl.\ Math.}}

\def\IJMPC{{\sl Int.\ J. Mod.\ Phys.\ C}}
\def\IJTP{{\sl Int.\ J. Theor.\ Phys.}}

\def\JSP{{\sl J. Stat.\ Phys.}}
\def\LAS{{\sl Los Alamos Science}}

\def\PD{{\sl Physica D}}

\def\PLB{{\sl Phys.\ Lett.\ B}}

\def\PR{{\sl Phys.\ Rev.}}
\def\PRA{{\sl Phys.\ Rev.\ A}}

\def\PRD{{\sl Phys.\ Rev.\ D}}
\def\PRE{{\sl Phys.\ Rev.\ E}}

\def\RMP{{\sl Rev.\ Mod.\ Phys.}}
\def\Sc{{\sl Science}}
\def\SPJETP{{\sl Sov.\ Phys.\ JETP}}

\def\ZP{{\sl Z. Physik}}

\def\dajm{\hbox{D. A. Meyer}}

\def\brosl{\hbox{B. Hasslacher}}
\def\bd{\hbox{\brosl\ and \dajm}}

\def\baxter{\hbox{R. J. Baxter}}
\def\feynman{\hbox{R. P. Feynman}}

\def\hfb{\hfil\break}

\catcode`@=11
\newskip\ttglue

   \font\ninerm=cmr9    \font\eightrm=cmr8   \font\sixrm=cmr6
  \font\ninebf=cmbx9   \font\eightbf=cmbx8  \font\sixbf=cmbx6
  \font\nineit=cmti9   \font\eightit=cmti8  
  \font\ninesl=cmsl9   \font\eightsl=cmsl8  
  \font\ninemi=cmmi9   \font\eightmi=cmmi8  \font\sixmi=cmmi6

\font\bigtenbf=cmr10 scaled\magstep2 

\def\ninepoint{\def\rm{\fam0\ninerm}%
  \textfont0=\ninerm \scriptfont0=\sixrm
  \textfont1=\ninemi \scriptfont1=\sixmi
  \textfont\itfam=\nineit  \def\it{\fam\itfam\nineit}%
  \textfont\slfam=\ninesl  \def\sl{\fam\slfam\ninesl}%
  \textfont\bffam=\ninebf  \scriptfont\bffam=\sixbf
    \def\bf{\fam\bffam\ninebf}%
  \tt \ttglue=.5em plus.25em minus.15em
  \normalbaselineskip=11pt
  \setbox\strutbox=\hbox{\vrule height8pt depth3pt width0pt}%
  \normalbaselines\rm}

\def\eightpoint{\def\rm{\fam0\eightrm}%
  \textfont0=\eightrm \scriptfont0=\sixrm
  \textfont1=\eightmi \scriptfont1=\sixmi
  \textfont\itfam=\eightit  \def\it{\fam\itfam\eightit}%
  \textfont\slfam=\eightsl  \def\sl{\fam\slfam\eightsl}%
  \textfont\bffam=\eightbf  \scriptfont\bffam=\sixbf
    \def\bf{\fam\bffam\eightbf}%
  \tt \ttglue=.5em plus.25em minus.15em
  \normalbaselineskip=9pt
  \setbox\strutbox=\hbox{\vrule height7pt depth2pt width0pt}%
  \normalbaselines\rm}

\def\sfootnote#1{\edef\@sf{\spacefactor\the\spacefactor}#1\@sf
      \insert\footins\bgroup\eightpoint
      \interlinepenalty100 \let\par=\endgraf
        \leftskip=0pt \rightskip=0pt
        \splittopskip=10pt plus 1pt minus 1pt \floatingpenalty=20000
        \parskip=0pt\smallskip\item{#1}\bgroup\strut\aftergroup\@foot\let\next}
\skip\footins=12pt plus 2pt minus 2pt
\dimen\footins=30pc

\def\ie{{\it i.e.}}

\magnification=1200
\input epsf.tex

\dimen0=\hsize \divide\dimen0 by 13 \dimendef\chasm=0
\dimen1=\hsize \advance\dimen1 by -\chasm \dimendef\usewidth=1
\dimen2=\usewidth \divide\dimen2 by 2 \dimendef\halfwidth=2
\dimen3=\usewidth \divide\dimen3 by 3 \dimendef\thirdwidth=3
\dimen4=\hsize \advance\dimen4 by -\halfwidth \dimendef\secondstart=4
\dimen5=\halfwidth \advance\dimen5 by -10pt \dimendef\indenthalfwidth=5
\dimen6=\thirdwidth \multiply\dimen6 by 2 \dimendef\twothirdswidth=6
\dimen7=\twothirdswidth \divide\dimen7 by 4 \dimendef\qttw=7
\dimen8=\qttw \divide\dimen8 by 4 \dimendef\qqttw=8
\dimen9=\qqttw \divide\dimen9 by 4 \dimendef\qqqttw=9

\parskip=0pt
\line{\hfil September 1996}
\line{\hfil{\it revised\/} March 1997}
\line{\hfil quant-ph/9703027}
\bigskip\bigskip
\centerline{\bf\bigtenbf QUANTUM LATTICE GASES AND THEIR INVARIANTS}
\vfill
\centerline{\bf David A. Meyer}
\bigskip 
\centerline{\sl Project in Geometry and Physics}
\centerline{\sl Department of Mathematics}
\centerline{\sl University of California/San Diego}
\centerline{\sl La Jolla, CA 92093-0112}
\centerline{dmeyer@chonji.ucsd.edu}
\vfill
\centerline{ABSTRACT}
\bigskip
\noindent The one particle sector of the simplest one dimensional 
quantum lattice gas automaton has been observed to simulate both the
(relativistic) Dirac and (nonrelativistic) Schr\"odinger equations, in
different continuum limits.  By analyzing the discrete analogues of
plane waves in this sector we find conserved quantities corresponding
to energy and momentum.  We show that the Klein paradox obtains so 
that in some regimes the model must be considered to be relativistic
and the negative energy modes interpreted as positive energy modes of
antiparticles.  With a formally similar approach---the Bethe 
{\it ansatz\/}---we find the evolution eigenfunctions in the 
{\it two\/} particle sector of the quantum lattice gas automaton and
conclude by discussing consequences of these calculations and their
extension to more particles, additional velocities, and higher 
dimensions.
\bigskip
\global\setbox1=\hbox{PACS numbers:\enspace}
\global\setbox2=\hbox{PACS numbers:}
\parindent=\wd1
\item{PACS numbers:}  03.65.-w,  
                      11.30.-j,  
                      02.70.-c,  
                      89.80.+h.  
\item{\hbox to \wd2{KEY\hfill WORDS:}}   
                      quantum lattice gas; conserved quantities;
                      exactly solvable model;
\item{}               quantum computation.
\vfill
\centerline{to appear in \IJMPC.}
\vfill
\hrule width2.0truein
\medskip
\noindent Expanded version of an invited talk presented at the Sixth 
International Conference on Discrete Models in Fluid Mechanics held at 
the Boston University Center for Computational Science, Boston, MA, 
26--28 August 1996.
\eject

\headline{\ninepoint\it QLGA and their invariants
          \hfil David A. Meyer}

\parskip=10pt
\parindent=20pt

\noindent{\bf 1.  Introduction}

\noindent The recent interest in quantum lattice gas automata (QLGA) 
is at least partly stimulated by the prospect of quantum computation.
Quantum computers are (as yet hypothetical) physical systems designed
and programmed to take advantage of the quantum mechanical phenomena
of superposition and interference in order to perform calculations 
more efficiently than is possible with a classical computer [1].  As
the size of semiconductor devices decreases, useful quantum coherence
first seems possible at the nanoscale [2] where the most plausible
computer architecture is that of a cellular automaton [3].  Thus it 
is natural to consider quantum computation with lattice gas automata
(LGA).

Just as classical LGA are more efficiently deployed to simulate 
physical processes like diffusion and fluid flow than for universal
computation [4], it seems likely that, as anticipated already by 
Feynman [5], the first useful quantum computation will be a QLGA
simulation of some quantum mechanical process.  Thus it is important
to understand the physics of QLGA:  their symmetries, their conserved
quantities, and their macroscopic/continuum limits.

In the next section we begin by reviewing the one particle sector of 
the simplest QLGA in one dimension.  Determining the discrete analogue
of plane waves, we find a complete set of invariants.  These conserved
quantities bear interpretation as energy and momentum, and enable the
exact computation of the evolution.

This model has been observed to simulate both the (nonrelativistic)
Schr\"odinger equation [6,7] and the (relativistic) Dirac equation
[8,9,10].  Although perhaps surprising at first, this is not 
inconsistent as the Dirac equation is well known to have a 
nonrelativistic limit [11] and, as Benzi and Succi have emphasized, 
numerical evolution of parabolic equations can be stabilized and 
localized by the use of artificially hyperbolic equations [6].  In 
Section 3 we use the physical interpretation of the conserved 
quantities in the model to distinguish relativistic and 
nonrelativistic regimes of the QLGA.  Specifically, we show that the 
Klein paradox obtains [12]:  a sufficiently discontinuous 
geometry/potential precludes interpretation of the QLGA as simulating 
a single nonrelativistic particle, compelling recognition of 
antiparticles in the model as well.

In Section 4 we extend the one particle model to a multiparticle 
QLGA.  Introduction of all the two particle and antiparticle 
scattering amplitudes makes the QLGA formally (almost) equivalent to 
the six vertex model [13].  The exact solvability of the six vertex
model leads us to expect a {\sl computable\/} complete set of 
invariants for the multiparticle QLGA.  Concentrating on the two 
particle sector of the model, we find these invariants explicitly
and explain their physical meaning.

In Section 5 we conclude with a brief discussion of our results and
their extension to more particles, additional velocities, and higher 
dimensions.

\medskip
\noindent{\bf 2.  The one particle sector}
\nobreak

\nobreak
\noindent A LGA should be envisioned as a collection of particles 
moving synchronously from vertex to vertex on a fixed graph (lattice) 
$L$:  At the beginning of each timestep each particle is located at 
some vertex and is labelled with a `velocity' indicating along which 
edge incident to that vertex it will move during the `advection' half 
of the timestep.  After moving along the designated edge to the next 
vertex, in the `scattering' half of the timestep the particles at each 
vertex interact according to some rule which assigns new `velocity' 
labels to each.  Here we will consider only the one dimensional 
lattice isomorphic to the integer lattice $\Z$.

A QLGA is a LGA for which the time evolution is unitary.  To make this
precise we must first identify the Hilbert space of the theory.  An 
orthonormal basis for the one particle sector $H_1$ of the Hilbert 
space $H$ of the one dimensional QLGA is given by 
$\{|x,\alpha\rangle\}$ (in the standard Dirac notation [14]), where 
$x \in L$ denotes position and $\alpha \in \{\pm 1\}$ denotes 
`velocity'.  At each time the state of the QLGA is described by a 
{\sl state vector\/} in $H_1$:
$$
\Psi(t) = \sum_{x,\alpha} \psi_{\alpha}(t,x) |x,\alpha\rangle,
$$
where the {\sl amplitudes\/} $\psi_{\alpha}(t,x) \in \C$ and the norm 
of $\Psi(t)$, as measured by the inner product on $H$, is:  
$$
1 = \sum_{x,\alpha} \overline{\psi_{\alpha}(t,x)} \psi_{\alpha}(t,x).
                                                            \eqno(2.1)
$$
The state vector evolves unitarily, \ie, $\Psi(t+1) = U \Psi(t)$, 
where $U$ is a unitary operator on $H$.  Since the evolution is 
unitary, the inner product is preserved and (2.1) holds for all times 
if it holds for one; this allows the interpretation of 
$\overline{\psi_{\alpha}(t,x)} \psi_{\alpha}(t,x)$ as the 
{\sl probability\/} that the particle be in the state 
$|x,\alpha\rangle$ at time $t$ [14,15].  The total probability is 
the most fundamental conserved quantity of a QLGA.

\moveright\secondstart\vtop to 0pt{\hsize=\halfwidth
\null
\vskip -1.75\baselineskip  
$$
\epsfxsize=\thirdwidth\epsfbox{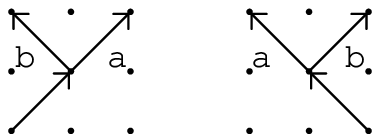}
$$
\vskip -0.5\baselineskip
\eightpoint{%
\noindent{\bf Figure 1}.  Spacetime diagrams of the one particle 
scattering amplitudes.  Time runs upward.
}}
\vskip-\baselineskip
\parshape=7
0pt \halfwidth
0pt \halfwidth
0pt \halfwidth
0pt \halfwidth
0pt \halfwidth
0pt \halfwidth
0pt \hsize
The `advection' followed by `scattering' evolution of the quantum
particle is implemented by having the basis vectors evolve as
$$
U|x,\alpha\rangle 
 = a|x+\alpha,\alpha\rangle + b|x+\alpha,-\alpha\rangle.    \eqno(2.2)
$$
That is, an initially right (left) moving quantum particle moves one 
lattice point to the right (left) and then continues in the same 
direction with amplitude $a$ or changes direction with amplitude $b$ 
as shown in Figure 1.  We showed in [10] that the most general
unitary evolution for a one dimensional QLGA with parity 
invariance%
\sfootnote*{{\it I.e.}, invariance under $x \to -x$; also called
            {\sl reflection\/} invariance.}
and particle speed 1 is described in the one particle sector by a 
unitary {\sl scattering matrix\/} 
$$
S := \pmatrix{b & a \cr
              a & b \cr
             }.                                             \eqno(2.3)
$$
$S$ may be parameterized by $a = \cos\theta$, $b = i\sin\theta$.
The model is then equivalent to Feynman's path integral for a Dirac
particle of mass $\theta$ [8], but one might ask for the origin of 
the evolution rule (2.2) and (2.3)---from what does the particle 
scatter?  One answer to this question will emerge at the end of the 
next section.

\moveright\secondstart\vtop to 0pt{\hsize=\halfwidth
\null
\vskip -3\baselineskip  
$$
\epsfxsize=\halfwidth\epsfbox{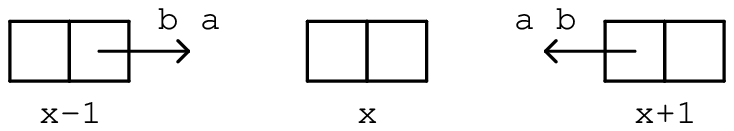}
$$
\eightpoint{%
\noindent{\bf Figure 2}.  Schematic diagram of a subspace of $H_1$.
Each box represents one dimension of the Hilbert space.  The arrows
indicate the amplitude flow into the two components at $x$; the 
labels $a$ and $b$ are the multiplicative factors to the corresponding 
boxes.
}}
\vskip-\baselineskip
\parshape=8
0pt \halfwidth
0pt \halfwidth
0pt \halfwidth
0pt \halfwidth
0pt \halfwidth
0pt \halfwidth
0pt \halfwidth
0pt \hsize
It is convenient to introduce the parity transformed scattering
matrix $\check S := PS$, where
$$
P := \pmatrix{0 & 1 \cr
              1 & 0 \cr
             }.
$$
Then (2.2) becomes
$$
\eqalignno{
\noalign{\smallskip}
U|x,\alpha\rangle 
 &= \sum_{\alpha'} \check S_{\alpha'\alpha} |x+\alpha,\alpha'\rangle.
                                                             &(2.4)\cr
}
$$
This is equivalent to a condition on the amplitudes:
$$
U \psi_{\alpha'}(x) 
 = \sum_{\alpha} \check S_{\alpha'\alpha} \psi_{\alpha}(x-\alpha),
                                                            \eqno(2.5)
$$
where the time argument is omitted.  Figure 2 illustrates the 
amplitude flow in $H_1$ described by (2.5).  The discrete analogues of 
{\sl plane waves\/} are {\sl wave functions\/} 
$\psi_{\alpha} : L \to \C$ which evolve by phase multiplication.  
Since $U$ is unitary, these are exactly its eigenvectors; they evolve
by multiplication by the corresponding norm 1 eigenvalues 
$e^{-i\omega}$.  For plane waves, therefore, (2.5) becomes
$$
e^{-i\omega} \psi_{\alpha'}(x) 
 = \sum_{\alpha} \check S_{\alpha'\alpha} \psi_{\alpha}(x-\alpha).
                                                            \eqno(2.6)
$$
Solving an equation like (2.6) is a problem we will face three times 
in this paper.  In each case we proceed the same way, by making an
appropriate {\it ansatz\/} for the form of a solution and then
constraining the parameters in the {\it ansatz\/} by imposing 
additional (matching) conditions.

As we explain in some detail in [16] for a QLGA which generalizes the
one defined by (2.2) and (2.3), the appropriate {\it ansatz\/} for 
(2.6) is
$$
\psi_{\alpha}^{(k)}(x-\alpha) = e^{-i\alpha k} \psi_{\alpha}^{(k)}(x),
                                                            \eqno(2.7)
$$
just as we would expect for a plane wave.  Plugging (2.7) into (2.6) 
gives
$$
e^{-i\omega} \psi_{\alpha'}(x) 
 = \sum_{\alpha} \check S_{\alpha'\alpha} e^{-i\alpha k} 
                 \psi_{\alpha}(x).                          \eqno(2.8)
$$
\eject
\noindent As the notation suggests, the $\psi_{\alpha}(x)$ are the 
components of a vector $\psi(x) \in \C^2$ (in this notation the state 
vector is $\Psi = \sum_x \psi(x) |x\rangle$); thus (2.8) can be 
rewritten as an eigenvalue problem:
$$
e^{-i\omega} \psi^{(k)}(x) = D(k) \psi^{(k)}(x),
$$
where
$$
D(k) := \pmatrix{ ae^{ik} & be^{-ik} \cr
                  be^{ik} & ae^{-ik} \cr
                }
$$
is unitary since $S$ is.  $D(k)$ summarizes the amplitude flow into 
the central pair of blocks in Figure 2.

\moveright\secondstart\vtop to 0pt{\hsize=\halfwidth
\null\vskip-2.5\baselineskip   
$$
\epsfxsize=\halfwidth\epsfbox{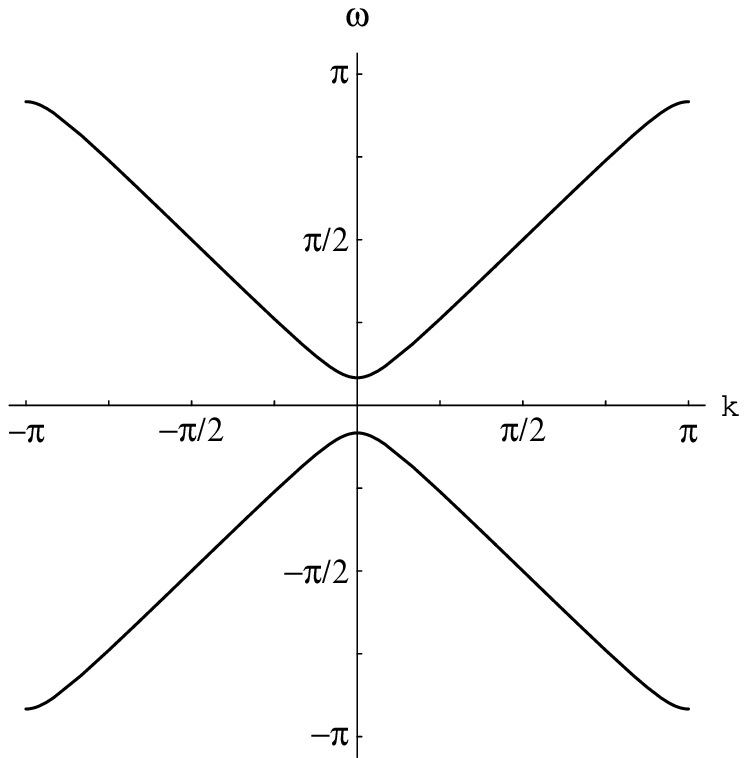}
$$
\eightpoint{%
\noindent{\bf Figure 3}.  The dispersion relation for 
$\theta = \pi/12$.  There are two branches and a frequency gap 
between $\theta$ and $-\theta$.
}}
\vskip-\baselineskip
\parshape=22
0pt \halfwidth
0pt \halfwidth
0pt \halfwidth
0pt \halfwidth
0pt \halfwidth
0pt \halfwidth
0pt \halfwidth
0pt \halfwidth
0pt \halfwidth
0pt \halfwidth
0pt \halfwidth
0pt \halfwidth
0pt \halfwidth
0pt \halfwidth
0pt \halfwidth
0pt \halfwidth
0pt \halfwidth
0pt \halfwidth
0pt \halfwidth
0pt \halfwidth
0pt \halfwidth
0pt \hsize
The characteristic equation for $D(k)$ is the {\sl dispersion 
relation\/}:
$$
\cos\omega = \cos\theta\cos k.                              \eqno(2.9)
$$
For pairs $(\omega,k)$ satisfying the dispersion relation, the plane
wave {\it ansatz\/} satisfies (2.6).  The dispersion relation (2.9) 
has two solutions $\pm\omega_k$ for each $k$, as shown in Figure 3.  
The eigenvalues $e^{\mp i\omega_k}$ have eigenvectors
$$
\psi^{(k,\pm)}(0) := \pmatrix{         -b e^{-ik}          \cr
                              a e^{ik} - e^{\mp i\omega_k} \cr
                             }.                             \eqno(2.10)
$$
The plane wave solutions to (2.6) are thus
$$
\eqalignno{
\noalign{\smallskip}
\psi^{(k,\pm)}(x) &= e^{ikx} \psi^{(k,\pm)}(0),             &(2.11)\cr
\noalign{\smallskip}
}
$$
where for positive $\omega_k$, positive (negative) $k$ gives a right 
(left) moving plane wave.  That the wave functions (2.11) are the 
discrete analogues of continuum plane waves is visible in Figures 4 
and 5 which show, respectively, the evolution of right moving 
$k = \pi/16$ and $k = \pi/8$ plane waves for $0 \le x \le 31$.

\midinsert
$$
\epsfxsize=\halfwidth\epsfbox{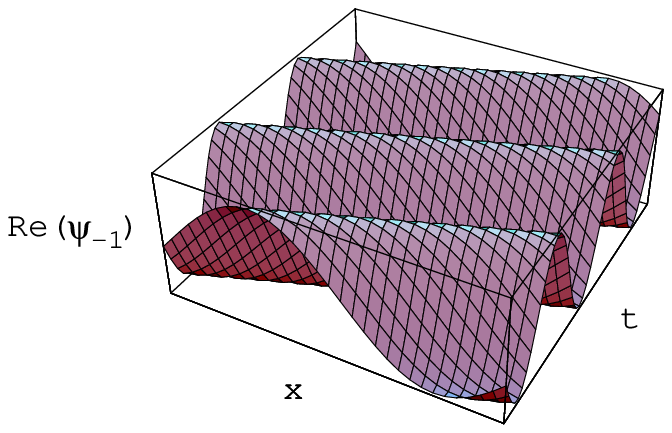}\hskip\chasm%
\epsfxsize=\halfwidth\epsfbox{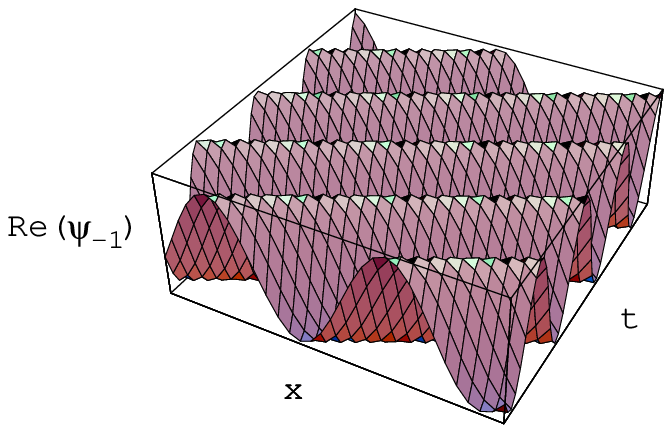}
$$
\hbox to\hsize{%
\vbox{\hsize=\halfwidth\eightpoint{%
\noindent{\bf Figure 4}.  Evolution of the $k = \pi/16$ right moving 
plane wave for the interval $0 \le x \le 31$ when $\theta = \pi/12$.
}}
\hfill%
\vbox{\hsize=\halfwidth\eightpoint{%
\noindent{\bf Figure 5}.  The same simulation for the $k = \pi/8$ 
right moving plane wave.  This wave has a higher frequency.
}}}
\endinsert

Just as in the continuum situation, therefore, $\omega$ and $k$ bear
interpretation as being proportional to energy and momentum, 
respectively [16].  Furthermore, they label a complete set of 
conserved quantities in terms of which the evolution can be exactly
computed:  Let
$$
|k,\epsilon\rangle
 := \sum_x \psi^{(k,\epsilon)}(0) e^{ikx} |x\rangle
$$
denote the plane wave state vectors in $H_1$.  Then (dealing with 
normalization in the usual way [8]) the set $\{|k,\epsilon\rangle\}$ 
is an orthonormal basis for $H_1$ [16].  Consider any state vector 
$\Psi \in H_1$:
$$
\eqalign{
\Psi 
 &= \sum_x \psi(x) |x\rangle                                       \cr
 &= \sum_{\vphantom{k}x} \psi(x) 
    \sum_{k,\epsilon} |k,\epsilon\rangle \langle k,\epsilon|x\rangle
                                                                   \cr
 &= \sum_{k,\epsilon} 
    \Bigl(\sum_{\vphantom{k}x} 
          \langle k,\epsilon|x\rangle \psi(x)
    \Bigr) 
    |k,\epsilon\rangle                                             \cr
}
$$
The parenthesized expression gives the components in the new 
$|k,\epsilon\rangle$ basis:
$$
\eqalignno{
\hat\psi_{\epsilon}(k) 
 &:= \sum_x \langle k,\epsilon|x\rangle \psi(x)                    \cr
 &\phantom{:}= \sum_x \Bigl(\sum_y \psi^{(k,\epsilon)}(0) e^{iky}
                                   |y\rangle
                      \Bigr)^{\dagger} |x\rangle \psi(x)           \cr
 &\phantom{:}= \bigl(\psi^{(k,\epsilon)}(0)\bigr)^{\dagger} 
               \sum_x \psi(x) e^{-ikx}                             \cr
 &\phantom{:}=: \bigl(\psi^{(k,\epsilon)}(0)\bigr)^{\dagger} 
                \hat\psi(k),                                       \cr
}
$$
where $\hat\psi(k)$ is the {\sl discrete Fourier transform\/} of 
$\psi(x)$.  The plane waves $|k,\epsilon\rangle$ evolve by phase 
multiplication so the probabilities 
$\overline{\hat\psi_{\epsilon}(k)}\hat\psi_{\epsilon}(k)$ are left
invariant by the evolution.  Since any initial state vector $\Psi(0)$
can be expressed in the plane wave basis this way, the existence of
these conserved quantities is equivalent to exact solvability for 
the one particle sector of the QLGA.

There is, however, a familiar problem with the physical interpretation
of these conserved quantities.  We have identified $\omega$ as being
proportional to energy, so for each $k$ there are plane waves with
either positive or negative energy.  We could restrict our state 
vectors to have nonzero amplitudes only in the directions 
$|k,+\rangle$, \ie, not allow any negative energy states; the 
conservation laws would preserve this condition.  This would, however, 
preclude highly localized states:  $\Psi = |x_0,+1\rangle$, to take an
extreme example, has nonzero amplitudes in {\sl all\/} the negative
energy directions $|k,-\rangle$.  Furthermore, introducing any
interaction into the model risks disturbing this symmetry and allowing 
the system access to the (apparently) unphysical negative energy 
states.  In the next section we will see exactly how this can happen 
and how we should reinterpret the model as a consequence.

\medskip
\noindent{\bf 3.  The Klein paradox}
\nobreak

\nobreak
\noindent A single electron moving ballistically in a nanoscale 
semiconducting device is a conceivable implementation of a QLGA.  
Alternatively, simulation of such a physical situation is a possible
application of a QLGA implemented in some other way.  In either case,
inhomogeneous potentials must be included in the model.  As explained
in [17,8,7,16], modifying the evolution rule by an inhomogeneous 
phase factor has the desired effect.  For a time independent potential, 
(2.5) is modified as
$$
U\psi_{\alpha'}(x) 
 = \sum_{\alpha} \check S_{\alpha'\alpha} e^{-i\phi(x-\alpha)} 
                 \psi_{\alpha}(x - \alpha),                 \eqno(3.1)
$$
where $\phi(x)$ is the potential.

Consider a simple potential step:
$$
\phi(x) = \cases{  0  & if $x \le 0$; \cr
                 \phi & if $x  >  0$, \cr
                }                                           \eqno(3.2)
$$
and let us study right moving plane waves scattering off the step.  
Using (3.1), the equation analogous to (2.6) is
$$
e^{-i\omega} \psi_{\alpha'}(x) 
 = \sum_{\alpha} \check S_{\alpha'\alpha} e^{-i\phi(x-\alpha)}
                 \psi_{\alpha}(x-\alpha).                   \eqno(3.3)
$$

For $x < 0$, inserting the potential (3.2) into (3.3) gives exactly
equation (2.6).  With an incident right moving plane wave we expect a
reflected left moving plane wave, so an appropriate {\it ansatz\/} is
$$
\psi^{(\omega)}(x) = \psi^{(k,+)}(x) + A \psi^{(-k,+)}(x)
\qquad 
\hbox{if}\quad x \le 0,                                     \eqno(3.4)
$$
where $(\omega,\pm k)$ satisfy the dispersion relation (2.9).

For $x > 1$, inserting the potential (3.2) into (3.3) gives
$$
e^{-i\omega} \psi_{\alpha'}(x) 
 = \sum_{\alpha} \check S_{\alpha'\alpha} e^{-i\phi} 
                 \psi_{\alpha}(x-a)
$$
which is equivalent to 
$$
e^{-i(\omega-\phi)} \psi_{\alpha'}(x)
 = \sum_{\alpha} \check S_{\alpha'\alpha} \psi_{\alpha}(x-a).
                                                            \eqno(3.5)
$$
Again (3.5) has the same form as (2.6) so we make the {\it ansatz}
$$
\psi^{(\omega)}(x) = B \psi^{(k',+)}(x)
\qquad
\hbox{if}\quad x \ge 1,                                     \eqno(3.6)
$$
for the transmitted amplitudes, where
$$
\cos(\omega-\phi) = \cos\theta \cos k'.                     \eqno(3.7)
$$

\topinsert
\null\vskip-4\baselineskip    
$$
\epsfxsize=\usewidth\epsfbox{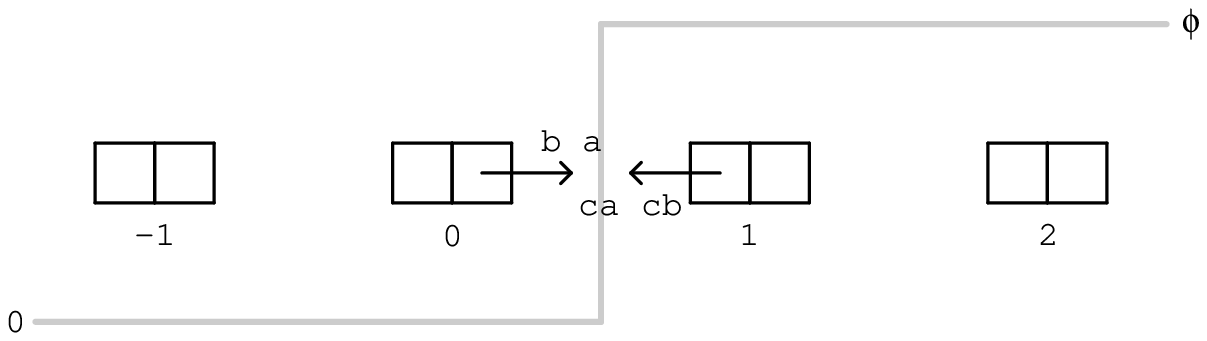}
$$
\vskip-2\baselineskip         
\eightpoint{%
{\narrower\noindent{\bf Figure 6}.  Schematic diagram of a subspace of
$H_1$ when there is a potential step of height $\phi$ between $x = 0$
and $x = 1$.  As in Figure 2, the boxes represent dimensions of the
Hilbert space and the labelled arrows indicate the amplitude flow 
(here $c := e^{-i\phi}$).  Equations (3.8) correspond to the 
statements that the left moving amplitude flow from $x = 1$ equals
what it would be were the {\it ansatz\/} for $x \le 0$ continued to
$x = 1$, and that the right moving amplitude flow from $x = 0$ equals
what it would be were the {\it ansatz\/} for $x \ge 1$ continued to
$x = 0$, respectively.\par}
}
\endinsert

As illustrated in Figure 6, at $x = 0$ and $x = 1$ (3.3) gives a pair 
of boundary conditions which constrain the two parameters $A$ and $B$:
$$
\eqalign{
e^{-i\omega} \psi_{\alpha'}^{(\omega)}(0) 
 &= \check S_{\alpha',-1} e^{-i\phi} \psi_{-1}^{(\omega)}(1) +
    \check S_{\alpha',+1}            \psi_{+1}^{(\omega)}(-1)      \cr
e^{-i\omega} \psi_{\alpha'}^{(\omega)}(1) 
 &= \check S_{\alpha',-1} e^{-i\phi} \psi_{-1}^{(\omega)}(2) +
    \check S_{\alpha',+1}            \psi_{+1}^{(\omega)}(0).      \cr
}
$$
These constraints imply that
$$
\eqalign{
B e^{-i\phi} \psi_{-1}^{(k',+)}(1) 
 &= 
\psi_{-1}^{(k,+)}(1) + A \psi_{-1}^{(-k,+)}(1)                     \cr
\psi_{+1}^{(k,+)}(0) + A \psi_{+1}^{(-k,+)}(0)
 &= 
B e^{-i\phi} \psi_{+1}^{(k',+)}(0),                                \cr 
}                                                           \eqno(3.8)
$$
after inserting the {\it ansatz\/} (3.4) and (3.6).  Using the 
formulae (2.10) and (2.11) for the plane waves and simplifying, the 
pair of equations (3.8) becomes
$$
\eqalign{
A         - B e^{-i\phi}        &= -1                              \cr
A e^{-ik} - B e^{-i(\phi - k')} &= -e^{ik}.                        \cr
}
$$
These equations can be solved for $A$ and $B$ to give
$$
\eqalign{
A &=         -{e^{ik'} - e^{ik} \over e^{ik'} - e^{-ik}}           \cr
B &= e^{i\phi}{e^{ik} - e^{-ik} \over e^{ik'} - e^{-ik}}.          \cr
}
$$
This justifies the {\it ansatz\/} of (3.4) and (3.6) as a solution to 
(3.3).

Notice that the solution $k'$ to (3.7) will not always be real, so
(3.6) and (3.7) will not always describe a plane wave.  For example, 
given $0 < \theta < \omega$, when the step is small, \ie, 
$0 < \phi < \omega - \theta$, the transmitted amplitudes form a plane
wave of lower frequency $\omega - \phi$ (and smaller wave number $k'$)
than the incident plane wave, but still lying on the upper branch of
the dispersion relation shown in Figure 3.  When the step is larger, 
\ie, $\omega - \theta < \phi < \omega + \theta$, $\omega - \phi$ lies
in the gap between the branches of the dispersion relation so (3.7) 
has purely imaginary solutions $k'$ and the only physical solution of 
the form (3.6) decays exponentially past the step.  

Figures 7 and 8 illustrate these situations on the interval 
$-128 \le x \le 128$ with $\theta = \pi/12$ for an incident right
moving plane wave with $\omega = \pi/6$ and 
$k = \arccos(\cos\omega/\cos\theta) \approx 0.459$.  In Figure 7 
the step has height $\pi/24$ and we have plotted the real part of 
$\psi_{-1}^{(\omega)}$.  This is the case 
$0 < \phi = \pi/24  < \omega - \theta = \pi/12$ so there is a
transmitted plane wave with 
$k = \arccos(\cos(\omega-\phi)/\cos\theta) \approx 0.296$.  In 
Figure 8 the step has height $\pi/8$; this is the case 
$\omega - \theta = \pi/12 < \phi = \pi/8 
                          < \omega + \theta = \pi/4$ so there is no
transmitted plane wave and the amplitude decays exponentially in $x$.

\topinsert
\null\vskip-\baselineskip
\vskip-\baselineskip
$$
\epsfxsize=\halfwidth\epsfbox{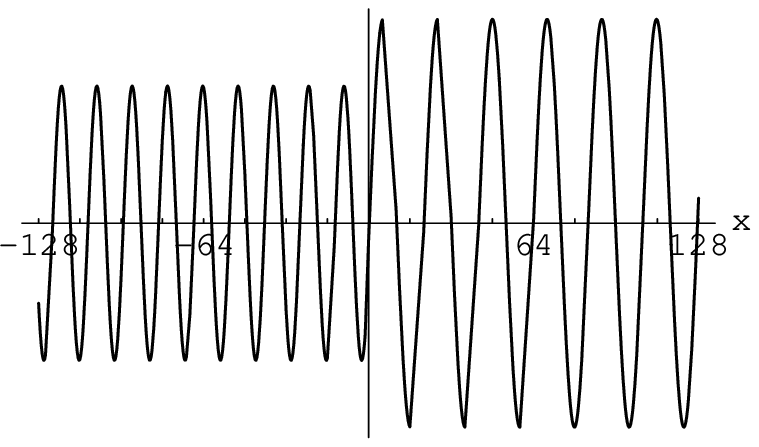}\hskip\chasm%
\epsfxsize=\halfwidth\epsfbox{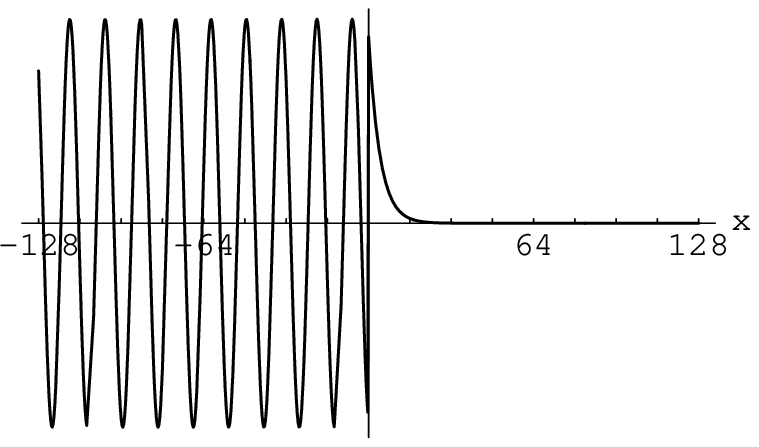}
$$
\hbox to\hsize{%
\vbox{\hsize=\halfwidth\eightpoint{%
\noindent{\bf Figure 7}.  The real part of the $\alpha = -1$ component
of an eigenfunction of $U$ with frequency $\omega = \pi/6$ for a step
of height $\pi/24$ with $\theta = \pi/12$ on the interval 
$-128 \le x \le 128$.  This wave function has wave numbers
aproximately 0.459 to the left of, and 0.296 to the right of the step.
}}
\hfill%
\vbox{\hsize=\halfwidth\eightpoint{%
\noindent{\bf Figure 8}.  The eigenfunction of $U$ in the same
situation as in Figure 7 except that in this case the step is 
taller---it has height $\pi/8$.  The incident right moving wave still 
has wave number approximately 0.459.  To the left of the step the wave 
function decays exponentially.
}}}
\endinsert

This much is exactly what we would expect in the nonrelativistic 
situation; and we would expect the transmitted amplitudes to simply 
decay exponentially faster as the height of the step is increased. 
Instead, we find that when the step is even larger, \ie, 
$\omega + \theta < \phi$, the transmitted amplitudes again form a 
plane wave, but now with frequency $\omega - \phi < -\theta < 0$.
Figure 9 illustrates this situation---the Klein paradox [12]---with a 
square well of depth $\phi = 7\pi/24$ and $\omega = \pi/6$ again, so 
that $\omega + \theta = \pi/4 < \phi = 7\pi/24$.   Here the 
transmitted plane wave has frequency $\omega - \phi = -\pi/8 < 0$ and 
hence negative energy.  This situation is out of the nonrelativistic 
regime; the large potential step localizes the particle too strongly, 
exciting negative energy states.

\moveright\secondstart\vtop to 0pt{\hsize=\halfwidth
\null\vskip-2.5\baselineskip
$$
\epsfxsize=\halfwidth\epsfbox{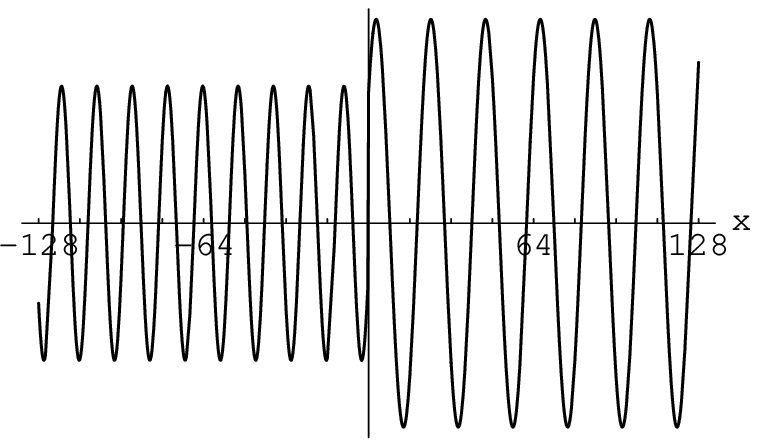}
$$
\eightpoint{%
\noindent{\bf Figure 9}.  The eigenfunction of $U$ in the same 
situation as in Figures 7 and 8, but now for an even taller step, of
height $7\pi/24$.  Rather than decaying even faster than in Figure 8,
the amplitude of the wave function oscillates to the right of the 
step, with wave number approximately $0.296$.
}}
\vskip-\baselineskip
\parshape=16
0pt \halfwidth
0pt \halfwidth
0pt \halfwidth
0pt \halfwidth
0pt \halfwidth
0pt \halfwidth
0pt \halfwidth
0pt \halfwidth
0pt \halfwidth
0pt \halfwidth
0pt \halfwidth
0pt \halfwidth
0pt \halfwidth
0pt \halfwidth
0pt \halfwidth
0pt \hsize
Just as in the continuum model for a relativistic quantum particle,
therefore, since we take negative energies to be unphysical, we are
forced to recognize the existence of {\sl anti\/}particles in the 
model, \ie, particles which are the time reverse $T$ (or charge 
conjugate $C$) of the particles with which we started.  The tokens 
representing the antiparticles must be the `empty' (lattice point, 
direction) pairs; that is, the single particle state 
$|x,\alpha\rangle$ should now be interpreted as a particle at 
$(x,\alpha)$ and antiparticles at all $(y,\beta) \not= (x,\alpha)$.  
Since the antiparticles are the time reverse of particles, a negative 
frequency corresponds to a {\sl positive\/} energy state for them; 
thus there are no unphysical negative energy states.  In this 
relativistic interpretation, the amplitude for the single particle to 
change direction is identified with the amplitude for it to scatter 
back from an oppositely moving antiparticle---answering the question 
asked in the previous section.  A complete reinterpretation of the 
model requires definition of the two antiparticle, and hence two 
particle, scattering amplitudes.  We do so, completing the definition 
of the QLGA, in the next section.

\medskip
\noindent{\bf 4.  The two particle sector}
\nobreak

\nobreak
\noindent In Section 2 we defined only the one particle sector of the
model; in [10] we showed that the most general parity invariant local
evolution rule which preserves particle number and has an exclusion
principle is defined by the unitary scattering matrix
$$
R = \bordermatrix{                             &
\scriptscriptstyle{\phantom{\nearrow\nwarrow}} &
\scriptscriptstyle{\phantom{\nearrow}\nwarrow} & 
\scriptscriptstyle{\nearrow\phantom{\nwarrow}} & 
\scriptscriptstyle{\nearrow\nwarrow}         \cr
\scriptscriptstyle{\phantom{\nwarrow\nearrow}} & d &   &   &   \cr
\scriptscriptstyle{\phantom{\nwarrow}\nearrow} &   & b & a &   \cr
\scriptscriptstyle{\nwarrow\phantom{\nearrow}} &   & a & b &   \cr
\scriptscriptstyle{\nwarrow\nearrow}           &   &   &   & f \cr
                 },                                         \eqno(4.1)
$$
where the central block is the scattering matrix $S$ (2.3) and 
unitarity requires $\bar dd = 1 = \bar ff$.  The phase $f$ is the 
amplitude for a pair of particles to scatter off one another, while
the phase $d$ is the amplitude for either a pair of empty states to
evolve to a pair of empty states or a pair of antiparticles to evolve
to a pair of antiparticles, depending on our interpretation of the 
model.  Figure 10 shows all the scattering rules implied by $R$; as in 
Figure 1 time runs upward so we indicate the presence of 
antiparticles by {\sl downward\/} pointing arrows.  This notation 
displays the (formal) equivalence of our one dimensional QLGA with 
two copies of the six vertex model [13].  The fact that two copies of
the six vertex model are required stems from the usual $\Z_2$ symmetry
in one dimensional LGA:  particles an even distance apart initially
cannot interact on the integer lattice [18,19].

\topinsert
\null\vskip-\baselineskip
$$
\epsfxsize=\usewidth\epsfbox{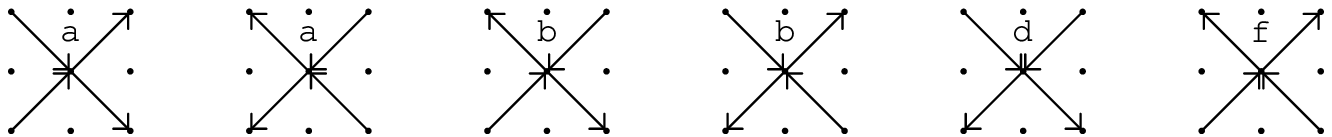}
$$
\vskip-\baselineskip
\eightpoint{%
{\narrower\noindent{\bf Figure 10}.  Spacetime diagrams of all the 
QLGA scattering amplitudes.  Time runs upward so antiparticles are 
represented by downward pointing arrows.\par} 
}
\endinsert

In the nonrelativistic interpretation it is most natural to set 
$d = 1$, while in the relativistic interpretation we must set 
$d = \bar f$ to maintain $CT$ invariance [20,19].  In either case, 
the evolution rule (2.2) for the basis vectors of the one particle 
sector of the Hilbert space gets multiplied by a factor of $d$ for 
each antiparticle/hole.  Since an overall phase is unobservable, we 
can always multiply $R$ by $\bar d$ to set $R_{11} = 1$ and avoid the 
normalization problem associated with an infinite number of factors
of $d$.  The one particle sector of the model is now completely 
defined and so supports the particle/antiparticle relativistic 
interpretation introduced at the end of the last section.

Furthermore, since particle number is preserved, only the relative 
phase between $S$ and $f$ has observable consequences, so the 
parameterization of $S$ in the previous sections can be maintained in 
each of the multiparticle sectors of the QLGA.  Since the 
inhomogeneous potentials considered in Section 3 are also implemented 
by a relative phase, this form for the scattering matrix $R$ suggests 
that similar calculations will determine the eigenfunctions of the 
evolution matrix for interacting particles.  In this section we 
demonstrate this explicitly for the two particle sector of the QLGA.

An orthonormal basis for the two particle sector $H_2 \subset H$ is 
given by the set
$\bigl\{|x_1,\alpha_1\rangle \otimes |x_2,\alpha_2\rangle 
 \bigm| 
 (x_1,\alpha_1) \not= (x_2,\alpha_2)\bigr\}$; this reflects
an exclusion principle in the model---the two particles cannot occupy
the same state---but not necessarily fermionic statistics.  A state 
vector for a pair of particles has the form
$$
\Psi(t) = \sum_{(x_1,\alpha_1) \not= (x_2,\alpha_2)}
           \psi_{\alpha_1\alpha_2}(t,x_1,x_2)
           |x_1,\alpha_1\rangle \otimes |x_2,\alpha_2\rangle
$$
and just as in the one particle sector $\Psi(t)$ must have norm 1:
$$
1 = \sum_{(x_1,\alpha_1) \not= (x_2,\alpha_2)}
     \overline{\psi_{\alpha_1\alpha_2}(t,x_1,x_2)}
     \psi_{\alpha_1\alpha_2}(t,x_1,x_2).
$$

The evolution of the two particle basis vectors implied by the
scattering matrix $R$ (4.1) divides into two cases.  For 
$x_1 + \alpha_1 \not= x_2 + \alpha_2$ the evolution rule (2.4) is 
simply applied to the two particles independently:
$$
U|x_1,\alpha_1\rangle \otimes |x_2,\alpha_2\rangle
 = \sum_{\alpha'_1,\alpha'_2} 
    \check S_{\alpha'_1\alpha^{\vphantom\prime}_1}
    \check S_{\alpha'_2\alpha^{\vphantom\prime}_2}
    |x^{\vphantom\prime}_1+\alpha^{\vphantom\prime}_1,
     \alpha'_1\rangle
    \otimes 
    |x^{\vphantom\prime}_2+\alpha^{\vphantom\prime}_2,
     \alpha'_2\rangle,                                      \eqno(4.2)
$$
while for $x_1 + \alpha_1 = x_2 + \alpha_2 = x$, there are two 
possible scattering events:
$$
\eqalignno{
U |x-1,+1\rangle \otimes |x+1,-1\rangle 
 &= f |x,+1\rangle \otimes |x,-1\rangle                            \cr
U |x+1,-1\rangle \otimes |x-1,+1\rangle
 &= f |x,-1\rangle \otimes |x,+1\rangle                            \cr
\noalign{\smallskip}
\noalign{\hbox{which can be summarized as}}
\noalign{\smallskip}
U |x-\alpha,\alpha\rangle \otimes |x+\alpha,-\alpha\rangle
 &= f |x,\alpha\rangle \otimes |x,-\alpha\rangle.            &(4.3)\cr
}
$$
The evolution rules (4.2) and (4.3) can be rewritten in terms of the
amplitudes:  For $x_1 \not= x_2$,
$$
\eqalignno{
U \psi_{\alpha'_1\alpha'_2}(x_1,x_2) 
 &= \sum_{\alpha_1,\alpha_2} 
     \check S_{\alpha'_1\alpha^{\vphantom\prime}_1}
     \check S_{\alpha'_2\alpha^{\vphantom\prime}_2}
     \psi_{\alpha^{\vphantom\prime}_1\alpha^{\vphantom\prime}_2}
      (x_1 - \alpha_1,x_2 - \alpha_2)                        &(4.4)\cr
\noalign{\hbox{and for $x_1 = x_2 = x$,}}
\noalign{\smallskip}
U \psi_{\alpha,-\alpha}(x,x)
 &= f \psi_{\alpha,-\alpha}(x - \alpha,x + \alpha).          &(4.5)\cr
}
$$
Just as Figures 2 and 6 illustrate the evolution rules (2.5) and 
(3.1), respectively, Figure 11 illustrates the amplitude flow on $H_2$
described by (4.4) and (4.5).

\topinsert
\null\vskip-2\baselineskip
$$
\epsfxsize=\twothirdswidth\epsfbox{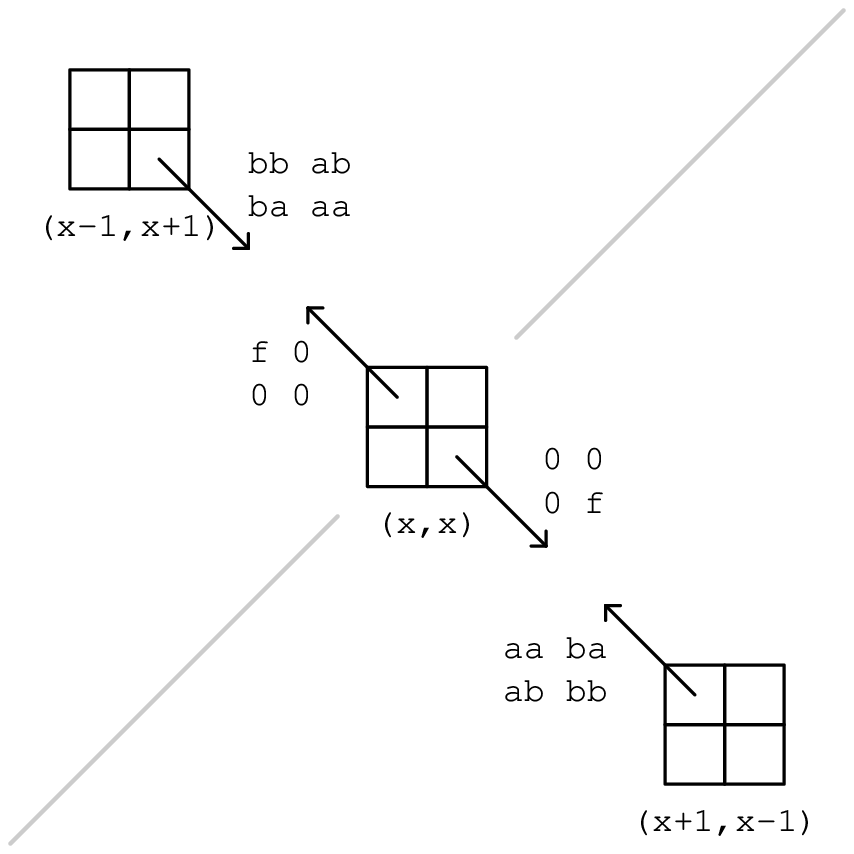}
$$
\vskip-\baselineskip
\eightpoint{%
{\narrower\noindent{\bf Figure 11}. Schematic diagram of a subspace
of $H_{2,{\rm int}}$.  The coordinates $x_1$ and $x_2$ of the two
particles label the sets of four boxes, each of which corresponds to
a pair of `velocities'.  The evolution rule (4.4) gives the amplitude 
flows shown pointing toward the diagonal (where the particles 
coincide), while (4.5) gives the amplitude flows shown pointing away 
from it.  In order for the {\it ansatz\/} (4.9) to satisfy (4.7) for 
$x_1 \not= x_2$, the amplitude flows away from the diagonal must 
satisfy (4.10); in order for it to satisfy (4.11) the amplitude flows 
toward the diagonal must satisfy (4.12).\par} 
}
\endinsert

Notice that we have implicitly assumed that the two particles are
distinguishable, as would be the case classically, and have 
interpreted the scattering events with amplitude $f$ to involve 
neither particle changing direction.  One might be concerned that 
(4.3) and (4.5) would have to be changed for indistinguishable quantum
particle scattering.  This is not the case when the particles have 
fermionic statistics, for example:  if the initial wave function is 
antisymmetric under the exchange of two particles, \ie,
$$
\psi_{\alpha_1\alpha_2} (x_1,x_2) 
 = -\psi_{\alpha_2\alpha_1} (x_2,x_1),                      \eqno(4.6)
$$
then the evolution rules (4.4) and (4.5) preserve this property.  We 
will consider this case in more detail at the end of this section.

At the beginning of this section we noted the close relation of our 
one dimensional QLGA with the six vertex model.  The six vertex model
is exactly solvable [21,13], which suggests that we should be able 
to find the analogue of plane waves in the two particle sector, namely
the two particle eigenfunctions of $U$.  The associated conserved
quantities will describe the physics of this sector.  For an 
eigenfunction, (4.4) becomes
$$
e^{-i\omega} \psi_{\alpha'_1\alpha'_2}(x_1,x_2) 
 = \sum_{\alpha_1,\alpha_2} 
    \check S_{\alpha'_1\alpha^{\vphantom\prime}_1}
    \check S_{\alpha'_2\alpha^{\vphantom\prime}_2}
    \psi_{\alpha^{\vphantom\prime}_1\alpha^{\vphantom\prime}_2}
     (x_1 - \alpha_1,x_2 - \alpha_2).                       \eqno(4.7)
$$
A straightforward {\it ansatz\/} for a solution to (4.7) is:
$$
\psi_{\alpha_1\alpha_2} (x_1,x_2)
 = \psi_{\alpha_1}^{(k_1,\epsilon_1)} (x_1) 
   \psi_{\alpha_2}^{(k_2,\epsilon_2)} (x_2),                \eqno(4.8)
$$
where $\psi^{(k_i,\epsilon_i)} =: \psi^{(i)}$ are the one particle 
plane waves we found in Section 2.  It is easy to verify that (4.8) 
solves (4.7):
$$
\eqalign{
U \psi_{\alpha'_1\alpha'_2} (x_1,x_2) 
 &= \sum_{\alpha_1,\alpha_2} 
     \check S_{\alpha'_1\alpha^{\vphantom\prime}_1}
     \check S_{\alpha'_2\alpha^{\vphantom\prime}_2}
     e^{-i\alpha_1k_1} 
     \psi_{\alpha^{\vphantom\prime}_1}^{(1)}(x_1)
     e^{-i\alpha_2k_2} 
     \psi_{\alpha^{\vphantom\prime}_2}^{(2)}(x_2)                  \cr
 &= \sum_{\alpha_1}
     \check S_{\alpha'_1\alpha^{\vphantom\prime}_1}
     e^{-i\alpha_1k_1}
     \psi_{\alpha^{\vphantom\prime}_1}^{(1)}(x_1)
     \sum_{\alpha_2} 
     \check S_{\alpha'_2\alpha^{\vphantom\prime}_2}
     e^{-i\alpha_2k_2} 
     \psi_{\alpha^{\vphantom\prime}_2}^{(2)}(x_2)                  \cr
 &= e^{-i\epsilon_1\omega_1} \psi_{\alpha'_1}^{(1)}(x_1)
    e^{-i\epsilon_2\omega_2} \psi_{\alpha'_2}^{(2)}(x_2)           \cr 
 &= e^{-i(\epsilon_1\omega_1 + \epsilon_2\omega_2)} 
    \psi_{\alpha'_1\alpha'_2} (x_1,x_2),                           \cr
}
$$
where the third equality follows from (2.8) since the pairs 
$(\epsilon_i\omega_i,k_i)$ satisfy the dispersion relation (2.9).  
Thus the {\it ansatz\/} (4.8) solves (4.7) for 
$\omega = \epsilon_1\omega_1 + \epsilon_2\omega_2$.

Now observe that the $\Z_2$ symmetry of the one dimensional QLGA 
partitions $H_2$ into two orthogonal subspaces:  
$H_2 = H_{2,{\rm int}} \oplus H_{2,{\rm free}}$, where 
$$
\eqalign{
H_{2,{\rm int}} 
 &:= \hbox{span}
      \bigl\{\, |x_1,\alpha_1\rangle \otimes |x_2,\alpha_2\rangle 
                \bigm|
                x_1 - x_2 \equiv 0 \hbox{\ (mod\ }2) \hbox{\ and\ }
                (x_1,\alpha_1) \not= (x_2,\alpha_2) \,
      \bigr\}                                                      \cr
H_{2,{\rm free}}
 &:= \hbox{span}
      \bigl\{\, |x_1,\alpha_1\rangle \otimes |x_2,\alpha_2\rangle 
                \bigm|
                x_1 - x_2 \equiv 1 \hbox{\ (mod\ }2) \,
      \bigr\}.                                                     \cr
}
$$
These subspaces evolve independently since $H_{2,{\rm int}}$ consists
of states in which the two particles can eventually interact while 
$H_{2,{\rm free}}$ consists of states in which they cannot, as is
implied by the form of (4.2) and (4.3).  In fact, the evolution of 
$H_{2,{\rm free}}$ is completely specified by (4.2) (or (4.4)) and 
depends only on the one particle scattering matrix $S$.  On this
subspace, therefore, (4.8) provides a complete solution for the 
eigenfunctions of $U$.

To find a solution for the interacting particles subspace 
$H_{2,{\rm int}}$, consider the two particle plane wave (4.8) 
`incident' from the $x_1 < x_2$ subspace of $H_{2,{\rm int}}$.  This
two particle plane wave `scatters' off the $x_1 = x_2$ subspace of 
$H_{2,{\rm int}}$ just as the right moving plane waves we considered 
in Section 3 scattered off the potential step, producing both a
`reflected' and a `transmitted' two particle plane wave, in the 
$x_1 < x_2$ and $x_1 > x_2$ subspaces, respectively.  Thus we make the
{\it Bethe ansatz\/} [22,21,13]:
$$
\psi_{\alpha_1\alpha_2}^{(\omega)} (x_1,x_2)
 = \cases{  \psi_{\alpha_1}^{(1)}(x_1) \psi_{\alpha_2}^{(2)}(x_2) +
          A \psi_{\alpha_2}^{(1)}(x_2) \psi_{\alpha_1}^{(2)}(x_1) 
          & if $x_1 < x_2$;                                        \cr
          B \psi_{\alpha_1}^{(1)}(x_1) \psi_{\alpha_2}^{(2)}(x_2) 
          & if $x_1 > x_2$,                                        \cr
         }                                                  \eqno(4.9)
$$
where $\omega := \epsilon_1\omega_1 + \epsilon_2\omega_2$, since all 
three two particle plane waves on the right hand side of (4.9) have 
the same two wave numbers $k_1$ and $k_2$ and the same frequency 
$\omega$.  For $|x_1 - x_2| \not= 2$, the free particle subspace 
computation together with linearity imply that (4.7) is satisfied by 
(4.9).  For $|x_1 - x_2| = 2$, (4.7) implies conditions on the 
extension of this {\it ansatz\/} to $x_1 = x_2 = x$:
$$
\eqalign{
e^{-i\omega} \psi_{\alpha'_1\alpha'_2}^{(\omega)} (x - 1,x + 1) 
 &= \sum_{(\alpha_1,\alpha_2) \not= (-1,+1)} 
     \check S_{\alpha'_1\alpha^{\vphantom\prime}_1}^{\vphantom|}
     \check S_{\alpha'_2\alpha^{\vphantom\prime}_2}^{\vphantom|}
     \psi_{\alpha^{\vphantom\prime}_1
           \alpha^{\vphantom\prime}_2
          }^{(\omega)} (x - 1 - \alpha_1,x + 1 - \alpha_2)         \cr
 &\qquad\qquad
    + \check S_{\alpha'_1,-1}^{\vphantom|}
      \check S_{\alpha'_2,+1}^{\vphantom|}
      \psi_{-1,+1}^{(\omega)} (x,x)                                \cr
e^{-i\omega} \psi_{\alpha'_1\alpha'_2}^{(\omega)} (x + 1,x - 1) 
 &= \sum_{(\alpha_1,\alpha_2) \not= (+1,-1)} 
     \check S_{\alpha'_1\alpha^{\vphantom\prime}_1}^{\vphantom|}
     \check S_{\alpha'_2\alpha^{\vphantom\prime}_2}^{\vphantom|}
     \psi_{\alpha^{\vphantom\prime}_1
           \alpha^{\vphantom\prime}_2
          }^{(\omega)} (x + 1 - \alpha_1,x - 1 - \alpha_2)         \cr
 &\qquad\qquad
    + \check S_{\alpha'_1,+1}^{\vphantom|}
      \check S_{\alpha'_2,-1}^{\vphantom|}
      \psi_{+1,-1}^{(\omega)} (x,x).                               \cr
}
$$
These conditions will be satisfied provided
$$
\eqalign{
\psi_{-1,+1}^{(\omega)} (x,x)
 &=  \psi_{-1}^{(1)} (x) \psi_{+1}^{(2)} (x) +
    A\psi_{+1}^{(1)} (x) \psi_{-1}^{(2)} (x)                       \cr
\psi_{+1,-1}^{(\omega)} (x,x)
 &= B\psi_{+1}^{(1)} (x) \psi_{-1}^{(2)} (x),                      \cr
}                                                           \eqno(4.10)
$$
\ie, provided the form of the {\it ansatz\/} (4.9) extends to the two
nonzero components of $\psi_{\alpha_1\alpha_2}^{(\omega)} (x,x)$ in 
this way.

Now the problem is to find $A,B \in \C$ such that (4.9) and (4.10)
define an eigenfunction of (4.5); that is, $\psi^{(\omega)}$ must also
satisfy
$$
e^{-i\omega} \psi_{\alpha,-\alpha}^{(\omega)} (x,x)
 = f \psi_{\alpha,-\alpha}^{(\omega)} (x - \alpha,x + \alpha). 
                                                           \eqno(4.11)
$$
This gives two constraints on $A$ and $B$:
$$
\eqalign{
&e^{-i\omega} \bigl[  \psi_{-1}^{(1)}(x) \psi_{+1}^{(2)}(x) +
                    A \psi_{+1}^{(1)}(x) \psi_{-1}^{(2)}(x)
              \bigr]
 = f B \psi_{-1}^{(1)}(x + 1) \psi_{+1}^{(2)}(x - 1)               \cr
&e^{-i\omega} B \psi_{+1}^{(1)}(x) \psi_{-1}^{(2)}(x) 
 = f \bigl[  \psi_{+1}^{(1)}(x - 1) \psi_{-1}^{(2)}(x + 1) +
           A \psi_{-1}^{(1)}(x + 1) \psi_{+1}^{(2)}(x - 1)
     \bigr].                                                       \cr
}                                                          \eqno(4.12)
$$
Inserting the form (2.11) of the one particle plane waves into (4.12)
and rearranging gives
$$
\eqalign{
A e^{-i\omega} \psi_{+-} - B e^{i(k_1 - k_2)} f \psi_{-+} 
 &= -e^{-i\omega} \psi_{-+}                                        \cr
A e^{i(k_1 - k_2)} f \psi_{-+} - B e^{-i\omega} \psi_{+-}
 &= -e^{-i(k_1 - k_2)} f \psi_{+-}                                 \cr
}                                                          \eqno(4.13)
$$
where 
$\psi_{\epsilon_1\epsilon_2} 
 := \psi_{\epsilon_1}^{(1)} (0) \psi_{\epsilon_2}^{(2)} (0)$.  Solving
(4.13) for $A$ and $B$ gives
$$
\eqalign{
A &= {\psi_{-+} \psi_{+-} (f^2 - e^{-2i\omega})
      \over
      (e^{-i\omega} \psi_{+-})^2 - (e^{i(k_1 - k_2)} f \psi_{-+})^2
     }                                                             \cr
B &= {e^{-i\omega} f \bigl[e^{-i(k_1 - k_2)} \psi_{+-}^2 -
                           e^{ i(k_1 - k_2)} \psi_{-+}^2 
                     \bigr]
      \over
      (e^{-i\omega} \psi_{+-})^2 - (e^{i(k_1 - k_2)} f \psi_{-+})^2
     }.                                                            \cr
}                                                          \eqno(4.14)
$$

Thus (4.9) and (4.10), with the coefficients $A$ and $B$ given by 
(4.14), define a set of eigenfunctions for the two interacting 
particles subspace $H_{2,{\rm int}}$.  Each is described by a pair of 
wave numbers $k_1$ and $k_2$, and a frequency
$\omega = \epsilon_1\omega_1 + \epsilon_2\omega_2$, where the pairs
$(\epsilon_i\omega_i,k_i)$ satisfy the dispersion relation (2.9).
When the two particles are apart, \ie, not at the same lattice point,
these eigenfunctions each put the particles into a superposition of 
two particle plane waves, with momenta proportional to $k_1$ and 
$k_2$.  Thus we should interpret these two particle eigenstates as
describing pairs of particles with distinct momenta before they
interact (\ie, arrive at the same lattice point), and the same 
distinct, but possibly permuted, momenta after they interact.  The
relative amplitudes of the permutations are $A$ and $B$ in (4.14).  To
understand their physical meaning, notice that $|A|^2 + |B|^2 = 1$.
Since $A$ and $B$ generically have both nonzero norms and phase 
angles, one may interpret $B$, for example, as describing the 
relative probability of transmission $|B|^2$ as well as a phase angle
shift $\Theta(k_1,k_2)$ between the incoming and scattered 
(transmitted) two particle plane waves.

When $A \not= 0$, a necessarily independent set of eigenfunctions for
the two interacting {\sl distinguishable\/} particles subspace 
$H_{2,{\rm int}}$ is obtained by the same calculation, only starting
with an `incident' two particle plane wave in the $x_1 > x_2$ 
subspace:
$$
\psi_{\alpha_1\alpha_2}^{(\omega)} (x_1,x_2)
 = \cases{B' \psi_{\alpha_1}^{(1)}(x_1) \psi_{\alpha_2}^{(2)}(x_2)
          & if $(x_1,\alpha_1) < (x_2,\alpha_2)$;                  \cr
              \psi_{\alpha_1}^{(1)}(x_1) \psi_{\alpha_2}^{(2)}(x_2) +
          A' \psi_{\alpha_2}^{(1)}(x_2)\psi_{\alpha_1}^{(2)}(x_1) 
          & if $(x_1,\alpha_1) > (x_2,\alpha_2)$,                  \cr
         }                                                 \eqno(4.15)
$$
where the ordering on pairs $(x_i,\alpha_i)$ is lexicographic and $A'$
and $B'$ are given by the formulae (4.14) for $A$ and $B$, but with
$k_1$ and $k_2$ exchanged (including the $k_1$ and $k_2$ in 
$\psi_{-+}$ and $\psi_{+-}$).

These eigenfunctions (4.15), together with our first set (4.9), form a
basis in terms of which antisymmetric wave functions (4.6) (and 
symmetric ones) can be expanded.  Since indistinguishable fermionic 
particles with a necessarily antisymmetric wave function are the most
plausible components of a QLGA quantum computer, however, it is 
perhaps more natural to work in the antisymmetric subspace of $H_2$ 
from the beginning.  Then the natural eigenfunction {\it ansatz\/} is
antisymmetric:
$$
\psi_{\alpha_1\alpha_2}^{(\omega)} (x_1,x_2)
 = \cases{\phantom{-}\psi_{\alpha_1}^{(1)}(x_1) 
                     \psi_{\alpha_2}^{(2)}(x_2) +
          A \psi_{\alpha_2}^{(1)}(x_2) \psi_{\alpha_1}^{(2)}(x_1) 
          & if $(x_1,\alpha_1) < (x_2,\alpha_2)$;                  \cr
                    -\psi_{\alpha_1}^{(1)}(x_1) 
                     \psi_{\alpha_2}^{(2)}(x_2) -
          A \psi_{\alpha_2}^{(1)}(x_2) \psi_{\alpha_1}^{(2)}(x_1) 
          & if $(x_1,\alpha_1) > (x_2,\alpha_2)$.                  \cr
         }                                                 \eqno(4.16)
$$
The {\it ansatz\/} (4.16) satisfies (4.7); the interaction condition
(4.11) imposes only a single constraint in this case:
$$
e^{-i\omega} \bigl[  \psi_{-1}^{(1)}(x) \psi_{+1}^{(2)}(x) +
                   A \psi_{+1}^{(1)}(x) \psi_{-1}^{(2)}(x)
             \bigr]
 =         f \bigl[- \psi_{-1}^{(1)}(x) \psi_{+1}^{(2)}(x) -
                   A \psi_{+1}^{(1)}(x) \psi_{-1}^{(2)}(x)
             \bigr].
$$
Inserting the form (2.11) of the one particle plane waves and solving
for $A$ gives
$$
A = -{e^{-i\omega} \psi_{-+} + f e^{-i(k_1 - k_2)} \psi_{+-}
      \over
      e^{-i\omega} \psi_{+-} + f e^{ i(k_1 - k_2)} \psi_{-+}
     }.                                                    \eqno(4.17)
$$
Thus (4.16) with the coefficient $A$ given by (4.17) defines a set of
antisymmetric eigenfunctions in the two interacting particles 
subspace; the eigenfunctions in the two free particles subspace are 
simply obtained by antisymmetrizing (4.8).

\medskip
\noindent{\bf 5.  Discussion}
\nobreak

\nobreak
\noindent We began our analysis in Section 2 by determining the 
eigenvalues and eigenfunctions---the phases $e^{-i\omega}$ and the 
plane waves $\psi^{(k,\epsilon)} \equiv |k,\epsilon\rangle$---of the
evolution operator $U$ on the one particle sector of the QLGA.  We 
found that the probability that a single particle with initial wave
function $\Psi(0)$ has wave number $k_0$ and frequency 
$\omega_0 = \epsilon_0 \arccos(\cos\theta\cos k_0)$ at time $t$ is an
invariant:  $|\langle\Psi(0)|k_0,\epsilon_0\rangle|^2$.  It is 
constant in time since each plane wave $|k,\epsilon\rangle$ evolves by
phase multiplication $e^{-i\omega t}$ over $t$ timesteps.  
Consequently, the expectation value of the wave number
$$
\langle k \rangle 
 := \sum_{k,\epsilon} k |\langle\Psi(0)|k,\epsilon\rangle|^2
$$
and of the frequency
$$
\langle \omega \rangle 
 := \sum_{k,\epsilon} \epsilon\omega_k 
                      |\langle\Psi(0)|k,\epsilon\rangle|^2
$$
are also invariants of the evolution.  Physically, of course, this
means that momentum and energy are conserved.

Identification of $k$ and $\omega$ as proportional to momentum and
energy, respectively, led to our interpretation in Section 3 of the
eigenfunctions of $U$ in the presence of a large step potential not as
negative energy modes of the particle but rather as positive energy
modes of antiparticles in the model.  Since there are no antiparticles 
in the (nonrelativistic) Schr\"odinger equation continuum limit of
this model, our analysis of invariants in the one particle sector of
the QLGA enables us to identify this as a parameter regime whose 
continuum limit must be described by the (relativistic) Dirac 
equation.

The eigenfunction {\it ansatz\/} and matching conditions we used for
the step potential forshadowed the Bethe {\it ansatz\/} approach we
applied in Section 4 to find the eigenvalues and eigenfunctions of $U$
in the two particle sector of the QLGA.  Just as the one particle
plane waves determine evolution invariants in the one particle sector,
the two particle eigenfunctions 
$|k_1,\epsilon_1;k_2,\epsilon_2\rangle := \psi^{(\omega)}$ defined by
(4.9) and (4.10), or (4.15), and (4.14) (or by (4.16) and (4.17)) 
define invariants in the two particle sector:  The probability that a 
pair of particles with initial wave function $\Psi(0)$ have wave 
numbers $k_i$ and frequencies 
$\omega_i = \epsilon_i \arccos(\cos\theta\cos k_i)$ at time $t$ is an
invariant:  $|\langle\Psi(0)|k_1,\epsilon_1;k_2,\epsilon_2\rangle|^2$.
Again, this is simply because the two particle eigenfunction evolves
by phase multiplication by $e^{-i\omega}$, where 
$\omega = \epsilon_1\omega_1 + \epsilon_2\omega_2$.

In fact, the Bethe {\it ansatz\/} extends to each of the multiparticle 
sectors of the QLGA [22,21,13] so the evolution of an arbitrary 
initial wave function can be computed exactly.  Of course, it had to 
be the case that the {\sl unitary\/} evolution operator $U$ has 
eigenfunctions which evolve by phase multiplication; what is 
remarkable, and what is meant by `exactly solvable', is that these 
eigenfunctions and eigenvalues can be effectively computed, just as we 
have done in the one and two particle sectors of the QLGA.

The most general, no less local, QLGA in one dimension includes
particles with `velocity' 0 [10].  We have determined the one 
particle plane waves for this model in [16] and it is natural to ask 
if the Bethe {\it ansatz\/} also solves this model exactly.  One 
expects this calculation to be possible, but more difficult:  $H_2$, 
for example, does not decompose into a direct sum of `interacting' and 
`free' subspaces as it does in the single speed case analyzed here.  
Similarly, Boghosian and Taylor have simulated the evolution of one 
particle plane waves in a two dimensional QLGA [7].  One would like 
to extend the Bethe {\it ansatz\/} to find multiparticle 
eigenfunctions in such a higher dimensional model, but should expect 
it to be even more difficult---there are very few higher dimensional
lattice models known to be exactly solvable [13,23].

\medskip
\noindent{\bf Acknowledgements}
\nobreak

\nobreak
\noindent It is a pleasure to thank Bruce Boghosian and the other 
organizers of the conference for inviting me to speak and for their
gracious hospitality.

\vfill\eject
\global\setbox1=\hbox{[00]\enspace}
\parindent=\wd1

\noindent{\bf References}
\bigskip

\parskip=0pt
\item{[1]}
D. P. DiVincenzo,
``Quantum computation'',
\Sc\ {\bf 270} (1995) 255--261;\hfb
I. L. Chuang, R. Laflamme, P. W. Shor and W. H. Zurek,
``Quantum computers, factoring, and decoherence'',
\Sc\ {\bf 270} (1995) 1633--1635;\hfb
A. Barenco and A. Ekert,
``Quantum computation'',
\APS\ {\bf 45} (1995) 1--12;\hfb
and references therein.

\item{[2]}
T. Ando, A. B. Fowler and F. Stem,
``Electronic properties of two-dimensional systems'',
\RMP\ {\bf 54} (1982) 437--672;\hfb
B. L. Altshuler, V. E. Kravtsov and I. V. Lerner,
``Distribution of mesoscopic fluctuations and relaxation processes
  in disordered conductors'',
in B. L. Altshuler, P. A. Lee and R. A. Webb, eds.,
{\sl Mesoscopic Phenomena in Solids\/}
(Amsterdam:  North-Holland 1991) 449--521;\hfb
D. K. Ferry and H. L. Grubin,
``Modeling of quantum transport in semiconductor devices'',
in H. Ehrenreich and F. Spaepen, eds.,
{\sl Solid State Physics:  Advances in Research and Applications\/}
{\bf 49} (1995) 283--448;\hfb
and references therein.

\item{[3]}
W. D. Hillis,
``New computer architectures and their relationship to physics or
  why computer science is no good'',
\IJTP\ {\bf 21} (1982) 255--262;\hfb
N. Margolus,
``Parallel quantum computation'',
in W. H. Zurek, ed.,
{\sl Complexity, Entropy, and the Physics of Information},
proceedings of the SFI Workshop, Santa Fe, NM, 
29 May--10 June 1989,
{\sl SFI Studies in the Sciences of Complexity} {\bf VIII}
(Redwood City, CA:  Addison-Wesley 1990) 273--287;\hfb
\brosl,
``Parallel billiards and monster systems'',
in N. Metropolis and G.-C. Rota, eds.,
{\sl A New Era in Computation}
(Cambridge:  MIT Press 1993) 53--65;\hfb
R. Mainieri,
``Design constraints for nanometer scale quantum computers'',
preprint (1993) LA-UR 93-4333, cond-mat/9410109;\hfb
M. Biafore,
``Cellular automata for nanometer-scale computation'',
\PD\ {\bf 70}\break
(1994) 415--433.

\item{[4]}
T. Toffoli, 
``Cellular automata as an alternative to (rather than an 
  approximation of) differential equations in modeling physics'',
\PD\ {\bf 10} (1984) 117--127;\hfb
\brosl,
``Discrete fluids'',
\LAS\ {\bf 15} (1988) 175--200, 211--217.

\item{[5]}
\feynman,
``Simulating physics with computers'',
\IJTP\ {\bf 21} (1982) 467--488.

\item{[6]}
S. Succi and R. Benzi,
``Lattice Boltzmann equation for quantum mechanics'',
\PD\ {\bf 69} (1993) 327--332;\hfb
S. Succi,
``Numerical solution of the the Schr\"odinger equation using 
  discrete kinetic theory'',
\PRE\ {\bf 53} (1996) 1969--1975.

\item{[7]}
B. M. Boghosian and W. Taylor, IV,
``A quantum lattice-gas model for the many-particle Schr\"odinger
  equation in $d$ dimensions'',
preprint (1996) BU-CCS-960401, PUPT-1615, quant-ph/9604035.

\item{[8]}
\feynman\ and A. R. Hibbs,
{\sl Quantum Mechanics and Path Integrals}
(New York:  McGraw-Hill 1965).

\vfill
\eject

\item{[9]}
I. Bialynicki-Birula,
``Weyl, Dirac, and Maxwell equations on a lattice as unitary 
  cellular automata'',
\PRD\ {\bf 49} (1994) 6920--6927.

\item{[10]}
\dajm,
``From quantum cellular automata to quantum lattice gases'',
\JSP\ {\bf 85} (1996) 551--574.

\item{[11]}
A. Messiah,
{\sl Quantum Mechanics}, Vol.\ II,
translated from the French by J. Potter
(Amsterdam:  North-Holland 1962) Chapter XX, Section V.

\item{[12]}
O. Klein,
``{\it Die Reflexion von Elektron an einem Potentialsprung nach der
       relativistischen Dynamik von Dirac}'',
\ZP\ {\bf 53} (1929) 157--165.

\item{[13]}
\baxter,
{\sl Exactly Solved Models in Statistical Mechanics\/}
(New York:  Academic Press 1982);\hfb
and references therein.

\item{[14]}
P. A. M. Dirac,
{\sl The Principles of Quantum Mechanics}, fourth edition
(Oxford:  Oxford University Press 1958).

\item{[15]}
H. Weyl,
{\sl The Theory of Groups and Quantum Mechanics},
translated from the second (revised) German edition by H. P. 
 Robertson
(New York:  Dover 1950).

\item{[16]}
\dajm,
``Quantum mechanics of lattice gas automata:  One particle
  plane waves and potentials'',
UCSD preprint (1996) quant-ph/9611005, to appear in \PRE.

\item{[17]}
G. V. Riazanov,
``The Feynman path integral for the Dirac equation'',
\SPJETP\ {\bf 6} (1958) 1107--1113.

\item{[18]}
G. Zanetti,
``Hydrodynamics of lattice gas automata'',
\PRA\ {\bf 40} (1989) 1539--1548;\hfb
Z. Cheng, J. L. Lebowitz and E. R. Speer,
``Microscopic shock structure in model particle systems:  The
  Boghosian-Levermore cellular automaton revisited'',
\CPAM\ {\bf XLIV} (1991) 971--979.

\item{[19]}
\bd,
``Lattice gases and exactly solvable models'',
\JSP\ {\bf 68} (1992) 575--590.

\item{[20]}
G. 't Hooft,
``A two-dimensional model with discrete general 
  coordinate-invariance'',
Duke University preprint (1989).

\item{[21]}
E. H. Lieb,
``Residual entropy of square ice'',
\PR\ {\bf 162} (1967) 162--172.

\item{[22]}
H. A. Bethe,
``{\it Zur Theorie der Metalle.  I.  Eigenwerte und Eigenfunktionen
       der linearen Atomkette}'',
\ZP\ {\bf 71} (1931) 205--226.

\item{[23]}
A. B. Zamolodchikov,
``Tetrahedra equations and integrable systems in 
  three-dimen\-sion\-al space'',
\SPJETP\ {\bf 52} (1980) 325--336;
``Tetrahedron equations and the relativistic $S$-matrix of 
  straight-strings in $2+1$-dimensions'',
\CMP\ {\bf 79} (1981) 489--505;\hfb
\baxter,
``On Zamolodchikov's solution of the tetrahedron equations'',
\CMP\ {\bf 88} (1983) 185--205;\hfb
J. M. Maillet and F. Nijhoff,
``Integrability for multidimensional lattice models'',
\PLB\ {\bf 224} (1989) 389--396.

\bye